\documentclass[prd,onecolumn,notitlepage,showpacs,preprintnumbers,amsmath,amssymb,nofootinbib,APS,10pt,superscriptaddress]{revtex4-1}

\usepackage{dcolumn}
\usepackage{bm}
\usepackage[spanish,english]{babel}
\usepackage[utf8]{inputenc}
\usepackage{amsmath,amssymb,amsfonts,latexsym,cancel}
\usepackage{graphicx}
\usepackage{color}
\usepackage{soul}
\usepackage{ulem}
\usepackage{hyperref}
\usepackage{slashed}
\usepackage{float}
\usepackage{xcolor}

\allowdisplaybreaks

\newcommand{\be}{\begin{equation}}
\newcommand{\ee}{\end{equation}}
\newcommand{\bea}{\begin{eqnarray}}
\newcommand{\eea}{\end{eqnarray}}

\begin{document}

\title{
Renormalization and a non-adiabatic vacuum choice in a radiation-dominated universe}

\author{Pau Beltr\'an-Palau}\email{pau.beltran@uv.es}
\affiliation{Departamento de Fisica Teorica and IFIC, Centro Mixto Universidad de Valencia-CSIC. Facultad de Fisica, Universidad de Valencia, Burjassot-46100, Valencia, Spain.}
\author{Sergi Nadal-Gisbert}\email{sergi.nadal@uv.es}
\affiliation{Departamento de Fisica Teorica and IFIC, Centro Mixto Universidad de Valencia-CSIC. Facultad de Fisica, Universidad de Valencia, Burjassot-46100, Valencia, Spain.}
\affiliation{Department of Physics and Astronomy, Louisiana State University, Baton Rouge, LA 70803, U.S.A.}
\author{Jos\'e Navarro-Salas}\email{jnavarro@ific.uv.es}
\affiliation{Departamento de Fisica Teorica and IFIC, Centro Mixto Universidad de Valencia-CSIC. Facultad de Fisica, Universidad de Valencia, Burjassot-46100, Valencia, Spain.}
\author{Silvia Pla}\email{silvia.pla@uv.es}
\affiliation{Departamento de Fisica Teorica and IFIC, Centro Mixto Universidad de Valencia-CSIC. Facultad de Fisica, Universidad de Valencia, Burjassot-46100, Valencia, Spain.}
\affiliation{Consortium for Fundamental Physics, School of Mathematics and Statistics,
Hicks Building, Hounsfield Road, Sheffield. S3 7RH United Kingdom. }

\begin{abstract}

Vacuum and particles can be naturally defined in the adiabatic regime of an expanding universe.
In general, however, there is no preferred choice of a vacuum state, unless  the spacetime background possesses special symmetries. In the absence of symmetries the standard viewpoint is to construct distinguished  adiabatic states permitting renormalizability of the stress-energy tensor. 
We study a special non-adiabatic vacuum for a massive scalar field in a radiation-dominated universe defined by imposing early-times conformal symmetry. We show that this state is consistent with renormalization, despite its ultraviolet behaviour is not the  conventional one due to the emergence of oscillatory terms.

 \end{abstract}

\date{\today}
\maketitle

\section{Introduction}

There are two basic issues in the theory of quantized fields in curved spacetime \cite{birrell-davies, Fulling, Wald94, parker-toms}. One is the explicit evaluation of the vacuum expectation values of product of fields, as the square field 
$\langle \phi^2 \rangle$ or the stress-energy tensor $\langle T_{\mu\nu}\rangle$. This requires to apply regularization and renormalization methods to tame the new ultraviolet divergences generated by spacetime curvature in a way consistent with general covariance. A second difficulty is to fix a preferred vacuum state.  The  canonical quantization approach exhibits    that the vacuum state is not unique in a general curved spacetime \cite{parker66}. 
Only for stationary spacetimes, or adiabatic regions in expanding universes, 
one can naturally find a privileged definition of the vacuum. A singular exception is for massless and conformally coupled fields in 
cosmological backgrounds. In this especial but important situation it is  possible to fix a natural (conformal) vacuum state \cite{parker66}.  
However, on a generic time-varying spacetime one cannot single out a preferred vacuum. For an incisive discussion, see \cite{Fulling, Fulling79}. This problem  has been recently reanalyzed in cosmological spacetimes in Ref. \cite{agullo-nelson-ashtekar},  emphasizing the role of the  so-called ultraviolet (UV) regularity condition. For earlier studies see also \cite{Anderson-Ad, AMM}.\\

The existence of underlying spacetime symmetries simplifies the above two issues, especially the  second.  In Minkowski space there exits an unambiguous construction of the vacuum state based on Poincaré invariance. The natural and Lorentz-invariant  splitting between positive and negative frequency modes serves to uniquely define a (Poincaré-invariant) vacuum and also to  identify the leading terms in the short-distance  behaviour of the two-point function. 
In de Sitter space things are slightly different. The underlying $SO(1,4)$ symmetry leads to an one-parameter family of invariant states \cite{Allen}, and one can select a unique vacuum (i.e., the Bunch-Davies vacuum \cite{BD-vacuum}) if  an ultraviolet condition is imposed (the so-called adiabatic condition \cite{parker-toms} for the modes). 
De Sitter space in planar coordinates $ds^2=dt^2 - e^{2Ht} d\vec x^2$ can be regarded as the limiting case of expanding universes with power-law expansion factors $a(t)\propto t^p$ with $p > 1$, as $p\to \infty$. Since those universes always behave adiabatically  at  early times, one can naturally privilege an UV regular initial vacuum state \cite{Weinberg, Abbot-Wise}. \\

In contrast, for an expansion factor of the form $a(t)\propto t^p$ with $p < 1$, the early time phase as $t\to 0$ encapsulates a non-adiabatic expansion. This makes more problematic to select a preferred initial vacuum. A prototype for this situation is a radiation-dominated universe, with $a(t)= a_0t^{1/2}$. The analysis of this special power-law expansion has  different motivations. It can be thought of as a natural  pre-inflationary phase, as it has been recently discussed in Ref. \cite{AndersonCarlson20}.  The vacuum in this pre-de Sitter space should smoothly evolve to an state that is approximately equivalent to the Bunch-Davies state for large $k$'s, but differs significantly from it at small $k$'s. 
This route can also be motivated by the results of loop quantum cosmology \cite{ashtekar-singh} or any other quantum gravity model completing the inflationary paradigm. Furthermore, from a non-inflationary perspective, the issue of how to define a preferred vacuum in a radiation-dominated universe is also of special relevance in general, and particularly in relation to the interesting and recent proposal of Ref. \cite{turok18, turok21}.\\

In the case of a radiation-dominated universe, one can select a preferred  vacuum state  for a scalar field by imposing conditions at $t\to 0$.  In this limiting situation, the mass term 
is irrelevant and the coupling $\xi$ to the vanishing scalar curvature is unimportant. Hence,  
the quantized scalar field behaves, at extremely  early times,  as a  conformal field. 
Imposing that the field modes behave  
as the exact modes of the conformally invariant theory selects a privileged vacuum\footnote{It is also interesting to remark the connection between the above field theoretical argument with the geometrical viewpoint underlying the Weyl curvature hypothesis advocated in \cite{Penrosebook}. The special nature of the (Bing-Bang) initial singularity, encapsulated in the assumption of a vanishing Weyl curvature, is what makes it possible to single out a preferred vacuum state in the quantized field theory.}$^{,}$\footnote{See Ref. \cite{particle-creation-76,GMM} for a vacuum characterization from the viewpoint 
of instantaneous Hamiltonian diagonalization and for $\xi=1/6$.}. 
However,  this vacuum is not an (infinite order) adiabatic state, like the Bunch-Davies vacuum or the adiabatic vacua in power-law expansions with $p>1$. This casts serious doubts on its consistency with renormalization. This work aims to show that the 
 vacuum defined this way is fully consistent with renormalization. For this, we mean that the vacuum expectation values of the renormalized stress-energy tensor 
 are finite and well-defined. This is a non-trivial issue since the analogous problem for a spin-$1/2$ field leads to a non-renormalizable stress-energy tensor \cite{Beltran-CPT}.\\

The paper is organized as follows. 
In Section \ref{sec:vacuum-choices} we discuss the issue of selecting an initial vacuum state for scalar fields propagating in isotropically expanding universes. The UV regularity condition and its role are analyzed in detail. In Section \ref{vacuum-RadiationDominated} we analyze a natural candidate for the initial vacuum state in a radiation-dominated universe. The choice of the state is based on an emergent conformal symmetry of the theory at very early times. Our analysis is focused on the ultraviolet behavior of the state. We show that it does not obey the conventional adiabatic condition. This is so because extra oscillatory terms appear in the large $k$ expansion. Nevertheless, we prove that the vacuum expectation values of the components of the renormalized stress-energy tensor are well-defined. Therefore, the proposed vacuum state seems to be a sensible physical state. In Section \ref{sec:conclusions} we state our conclusions and comment on further implications of our results. We use units such that $\hbar=1=c$. Furthermore, we follow the signature conventions of Refs. \cite{birrell-davies, parker-toms}.    
\\

\section{The choice of the initial vacuum state}
\label{sec:vacuum-choices}

In this section, we briefly discuss the question of how to select a preferred vacuum state in an expanding universe. We illustrate this issue 
with relevant examples, including de Sitter space. It will pave the way to introduce 
the (adiabatic) renormalization scheme and the UV regularity condition to be imposed on any vacuum state candidate. \\

Let us consider a spatially flat Friedmann-Lemaître-Robertson-Walker (FLRW) spacetime with metric $ds^2= dt^2 -a^2(t)d\vec x^2$.  The field equation for a massive scalar field is 
\be
(\Box + m^2 + \xi R)\phi=0\label{KG}\, ,
\ee
with $R=6(\dot{a}^{2} / a^{2}+\ddot{a} / a)$ and where $\xi$ is a dimensionless constant.  Due to spatial homogeneity, we can perform a generic mode expansion of the scalar field as follows 
    \be
               \phi(t,\vec{x})=\int \frac{d^3k}{\sqrt{2(2\pi)^3a(t)^3}}\Big(A_{\vec{k}}e^{i\vec{k}\vec{x}}h_{k}(t)+A^\dagger_{\vec{k}}e^{-i\vec{k}\vec{x}}h^*_{k}(t)\Big)\,, \label{h-mode-expansion}
               \ee
where $A_{\vec{k}}$ and $A_{\vec{k}}^\dagger$ are the annihilation and creation operators satisfying the conventional commutation relations. 
The time-dependent part of the mode functions $h_k(t)$ satisfies 
\be \ddot h_k +[\omega_k(t)^2 +\sigma(t)]h_k=0\,,\label{motion-h}\ee
with $\omega^2_k=\frac{k^2}{a^2}+m^2$, $k^2=\vec{k}^2$ and $\sigma=(6\xi-\frac{3}{4})(\frac{\dot a}{a})^2+(6\xi-\frac{3}{2})\frac{\ddot{a}}{a}$, together with the  normalization Wronskian condition \be \label{Wronskian}
         \dot{h}_k^*h_k-h_k^*\dot{h}_k=2i\,.
               \ee               

For the de Sitter space $a(t)= e^{Ht}$ the  mode equation   becomes        
\be
\ddot h_k(t) + \Big[k^2e^{-2H t}+m^2+H^2(12 \xi-\frac{9}{4})\Big]h_k(t)=0.
\ee
This differential equation is a Bessel's equation
in the variable $v= kH^{-1}e^{-Ht}$, with index $\nu^2 = 9/4 -
(m^2+ 12\xi H^2)/H^2$,
and its solution can be written in terms of Hankel functions $H^{(1)}_\nu(v)$ and $H^{(2)}_\nu(v)$
\be h_k(t) =
\sqrt{\frac{\pi}{2 H}}[E_kH^{(1)}_{\nu}(kH^{-1}e^{-Ht})+F_kH^{(2)}_{\nu}(kH^{-1}e^{-Ht})] \ , 
\ee
where the dimensionless constants $E_k$ and $F_k$ obey the normalization condition $|E_k|^2-|F_k|^2=1$. We have assumed for simplicity $\nu^2 >0$. In order to fix the coefficients $E_k$ and $F_k$ we can take advantage of  de Sitter symmetries. It is not difficult to see \cite{Higuchi87,parker-toms} that de Sitter invariance requires that the coefficients $E_k$ and $F_k$ are independent of $k$. Therefore, one gets a one-parameter family of de Sitter invariant vacuum choices \cite{Allen,LesHouches} (i.e., the well-known alpha vacua). \\

An additional requirement that 
the vacuum should satisfy is that it has to be 
 ultraviolet (UV) regular. 
This can be easily understood by requiring that the short distance behavior of the two-point function must be similar to that found in Minkowski space. In terms of field modes this means that, for very large $k$, 
 the field modes must behave as
\be \label{conformalmodes}h_k(t)\sim \frac{1}{\sqrt{W_k(t)}}e^{-i\int^t W_k(t') dt'}  \ , \ee 
where $W_k(t) = \omega_k(t) + \cdots$ (see next subsection for details).
 In de Sitter space this implies \be h_k(t)\sim
\frac{1}{\sqrt{ke^{-Ht}}}e^{i (k H^{-1}e^{-Ht})} \ . \ee
This fixes univocally the numeric coefficients $ E= 1$ and $F=0$ and defines  the so-called Bunch-Davies vacuum. 
The two natural requirements that we have imposed above, i.e. the fact that the vacuum has to respect the symmetries of the spacetime and the UV regularity condition, 
allowed us to fix unambiguously the  vacuum in de Sitter spacetime.\\ 

However, in general spatially flat FLRW universes (which are invariant only under spatial rotations and translations), selecting a preferred vacuum is more problematic. Nevertheless, for expansions of the form $a(t)=a_0t^{p}$ with $p>1$ we can still choose a preferred vacuum on the basis of the large $k$ expansion (\ref{conformalmodes}).   This can be deduced from the specific form of mode equation (\ref{motion-h})
\be
\ddot h_k  +[m^2 + \frac{k^2}{a_0^2t^{2p}} +  (6\xi-\frac{3}{4})\frac{p^2}{t^2}+(6\xi-\frac{3}{2})\frac{p(p-1)}{t^2}]h_k=0 \label{motion-h2N} \ . \ee
We observe that, for $t \to 0$, the physical momentum of a given mode $k^2/a^2(t)$ largely surpasses the background scale $\sigma(t)= (6\xi-\frac{3}{4})\frac{p^2}{t^2}+(6\xi-\frac{3}{2})\frac{p(p-1)}{t^2}$.
Therefore, we can heuristically say that the  
expansion of the spacetime is adiabatic with respect to the natural scale of  field modes at early times.
The requirement (\ref{conformalmodes}) for any $k$ at early times $t\to 0$ univocally determines a vacuum state. It is useful to illustrate the above with a simple  example. For $m^2=0$ and $\xi=0$ one can analytically solve the mode equation in terms of the 
conformal time $\tau=\int a^{-1}dt$  and fix the  vacuum by demanding \eqref{conformalmodes} when $t\to 0$ ($\tau \to -\infty$). The adiabatic solution for any $k$ is given by \cite{Ford-Parker77,Weinberg,Abbot-Wise}

\be \label{solutionFPWN}h_k(\tau) = e^{\frac{i\pi(3+2\nu)}{4}}\sqrt{\frac{-\pi \tau a(\tau)}{2}}  H_{\nu}^{(1)}(- k \tau ) \ee
with $\nu=\frac{3p-1}{2(p-1)}$. It is easy to check that the large $k$ expansion of (\ref{solutionFPWN}) agrees with (\ref{conformalmodes}) exactly. The same is true for massive fields. We remark that we are only forced to impose the UV regularity condition  
for $k\to \infty$. This leaves unspecified the infrared characterization of the vacuum.
It is conventional in cosmology \cite{Weinberg} to impose (\ref{conformalmodes}) for every $k$, which singles out a particular vacuum state.  
As stated in the above examples, the relevance of the UV regularity condition motivates a more detailed analysis.
 \\

\subsection{UV regularity condition}

As we have argued previously, an unavoidable requirement that any suitable vacuum state must meet is that it has to be ultraviolet (UV) regular. It means that the formal vacuum expectation values of relevant observables have to possess the same local divergences as in Minkowski spacetime as well as some additional divergent terms due to the curved background. This becomes necessary to guarantee the existence of finite vacuum expectation values after renormalization. 
 For quantum states in FLRW spacetimes this criterion can be implemented by the {\it adiabatic condition} \cite{parker-toms} [Section 3.1] (see also \cite {agullo-nelson-ashtekar} and references therein). Let us analyze it in terms of the two-point function. From the mode expansion of the scalar field \eqref{h-mode-expansion} we can easily compute the formal vacuum expectation value of the two point-function as 
 \be                \langle \phi^2\rangle =\frac{1}{(2\pi)^2a(t)^3}\int_{0}^{\infty} dk k^2 |h_{k}(t)|^2 \label{phi2}\, .                \ee
This quantity is ultraviolet divergent and has to be renormalized to obtain a finite, physical value. To this end, one has to be able to identify first its UV divergences.  
In general FLRW spacetimes, we have a very convenient tool to determine them univocally
: {\it the adiabatic expansion of the modes $h_k(t)$}. For scalar fields, it is based on the WKB ansatz, namely \cite{parker-fulling} (see also \cite{birrell-davies, Fulling, parker-toms})

\be \label{adiabatic-expansionN}h_k^{\rm Ad}(t) \sim \frac{1}{\sqrt{W_k}}e^{- i\int^t W_k(t')dt'}\,
\ee
where the function $W_k(t)$ admits an adiabatic expansion in terms of the derivatives of $a(t)$
\be \label{adiabaticexpansionN}W_k = \omega_k + \omega_k^{(1)}+ \omega_k^{(2)} + \omega_k^{(3)}+ \omega_k^{(4)} + \cdots   \ . \ee
The coefficient $\omega_k^{(n)}$ depends on derivatives of $a(t)$ up to and including the order $n$. We note here that the square of frequency scale of the background $\sigma(t)$ is a function of adiabatic order two.  
 The leading order of the expansion is $\omega_k^{(0)}\equiv \omega_k=\sqrt{k^2/a^2+m^2}$ and the next-to-leading orders are obtained, by systematic iteration, from the relation
\be \label{ad_eq}
W_k^2=\omega_k^2+\sigma + \frac{3}{4}\frac{\dot W_k^2}{W^2_k}-\frac{1}{2}\frac{\ddot W_k}{W_k} \ ,
\ee
derived from the mode equation \eqref{motion-h}. Inserting the adiabatic expansion in the equation above, and grouping terms with the same adiabatic order, it is possible to obtain the $n$th coefficient from the lower ones once the leading term is defined. It can be proved that the terms with odd adiabatic order are zero, i.e., $\omega_k^{(2n+1)}=0$. The first next-to-leading order terms can be found, for example, in \cite{parker-toms}. With the expansion \eqref{adiabatic-expansionN}, we can easily expand $|h_k|^2$ in terms of the adiabatic function $W_k$ as   \be \label{ad-exp-hk2}|h_k^{\rm Ad}|^2\sim (W_k^{-1})^{(0)}+ (W_k^{-1})^{(2)}+(W_k^{-1})^{(4)}+\cdots\ee
that, for large momentum $k$ reads
 \bea
 (W_{k\to \infty}^{-1})^{(0)}&\sim&\frac{a}{k}-\frac{m^2a^3}{2k^3}+\frac{3m^4a^5}{8k^5}-\frac{5 m^6 a^7}{16 k^7}+ \mathcal{O}(k^{-9})\,,\\
 (W_{k\to \infty}^{-1})^{(2)}&\sim&
 -\frac{(\xi-\tfrac{1}{6})Ra^3}{2k^3}+\frac{c^{(2)}_5}{k^5}+ \frac{c^{(2)}_7}{k^7}+\mathcal{O}(k^{-9}) \label{adiabatic-largek-2}\,,\\
 (W_{k\to \infty}^{-1})^{(4)}&\sim&
 +\frac{c^{(4)}_5}{k^5} + \frac{c^{(4)}_7}{k^7}+\mathcal{O}(k^{-9})\,, \label{adiabatic-largek-4}
 \eea
 where the coefficients $c_n^{(i)}$ depend on the scale factor $a$, its derivatives, the mass of the scalar field $m$ and the scalar coupling $\xi$.  From this expansion, we immediately see that the divergent UV terms are captured in the zeroth and second order of the adiabatic expansion when integrated in \eqref{phi2}. \\
 
 The UV regularity condition required for any physical vacuum in a FLRW universe becomes now straightforward. One should require that the large momentum expansion of $|h_k|^2$ has to agree with the large momentum behavior of the adiabatic expansion \eqref{ad-exp-hk2} up to a given adiabatic order. 
 As argued above, with the adiabatic expansion, we capture all potential UV divergences in the leading orders. Therefore, we can  {\it renormalize} the two-point function by subtracting up to and including the adiabatic order two. 
 This procedure results in (some terms can be integrated exactly)
                \be \label{TPFfinalN}\langle\phi^2\rangle_{\textrm{ren}}=  \frac{1}{4 \pi^2 a^3}  \int_0^{\infty} d k k^2  \,  \left[|h_k|^2- \frac{1}{\omega_k} - \frac{(\frac{1}{6} -\xi)R}{2\omega_k^3}\right] - \frac{R}{288 \pi^2} \, .
\ee

  The same procedure applies to compute other relevant observables, such as the stress-energy tensor $\langle T_{\mu \nu}\rangle$.
The number of subtractions is determined by the scaling dimension of the observable. For the stress-energy tensor, this results in subtracting up to and including the fourth adiabatic order. This renormalization scheme 
is known as the adiabatic regularization method. The method was proposed in \cite{parker-fulling} for scalar fields and further improved in \cite{Anderson-Parker}. It has also been extended to fermions in \cite{adiabaticfermions}. It is used routinely in many works of quantum field theory in cosmological spacetimes, and it has also been reviewed in \cite{birrell-davies, Fulling, parker-toms}. 
The adiabatic regularization method can be regarded as an upgraded form of the generic DeWitt-Schwinger renormalization scheme when the spacetime background possesses the $3$-translational symmetry of the FLRW universes \cite{beltran-nadal}. Therefore, the subtracting terms are univocally fixed by the divergent terms of the DeWitt-Schwinger expansion, which respects manifest covariance for a general spacetime.\\

In summary, the {\it UV (adiabatic) regularity condition} is usually required for any quantum state in a FLRW spacetime. This means that the large momentum expansion of its associated modes  coincides exactly with the adiabatic expansion for large $k$ (at least 
up to and including the leading term of the 4th adiabatic order, e.g., $\mathcal{O}(k^{-5})$ for $|h_k|^2$) to ensure the renormalizability of the stress-energy tensor. The criterion guarantees the existence of a finite vacuum expectation value for the main physical observables. When the large momentum expansion coincides with the adiabatic expansion at any order we say that the vacuum is of infinite adiabatic order. This is the case of the preferred vacua analyzed in the first part of this section.\\
 
 The problem of constructing  a preferred vacuum for $a(t)=a_0 t^{p}$ with $p<1$ is more involved. As remarked in the introduction, this is so because at early times the expansion is largely non-adiabatic. In the following section, especially for the radiation dominated universe $a(t)=a_0 t^{1/2}$, we investigate a natural vacuum state that does not satisfies the conventional UV regularity condition.

\section{Renormalizability of the early-times non-adiabatic vacuum in a radiation dominated universe} \label{vacuum-RadiationDominated}

Let us focus now on a radiation-dominated universe, described by a scale factor of the form $a(t)=a_0 t^{1/2}$. The differential equation for the scalar field modes \eqref{motion-h} reads 
\be
\frac{d^{2} h_{k}}{d t^{2}}+\left(\frac{k^{2}}{a_{0}^{2} t}+m^{2}+\frac{3}{16 t^{2}}\right) h_{k}=0\, . \label{rad-h}
\ee
In the limit $t\to 0$, the expansion of this spacetime is not adiabatic 
with respect to the scale of the field modes, unlike the cases studied in the previous section. One can see this by analyzing this limit in the differential equation, %
where the term $3/(16t^2)$ (the background scale) dominates with respect to $k^2/(a_0^2t)$ (the physical momentum squared). Therefore one cannot easily define at early times an adiabatic vacuum in this universe. However, as time evolves the expansion of the universe slows down and one can naturally define a late-times (infinite order) adiabatic vacuum \cite{Birrell-Whitakker}. 
Let us briefly introduce the details of this choice for pedagogical reasons.
The solution to equation \eqref{rad-h} can be written in terms of the Whittaker functions $W_{\kappa, \mu}(z)$ \cite{mat-functions} as
     \be
     \label{solutionM}
      h_{k}=m^{-1/ 2}\Big[b_{k} W_{-i \lambda, \frac14}(i \bar t\,)+c_{k} W_{i \lambda, \frac14}(-i \bar t\,)\Big],
     \ee
     where  $\bar t = 2mt$, $\lambda=\frac{k^2}{2a_0^2m}$ , and $b_k$ and $c_k$ are dimensionless constants. From the properties of the Whittaker functions one can see that, at late times, the first term of this expression behaves as $e^{-imt}$ while the second one goes as $e^{imt}$. Since in the limit $t\to\infty$ the expansion of the universe tends to be adiabatic, the solution must behave 
     as the WKB adiabatic expansion  \eqref{adiabatic-expansionN}. 
     The constants are fixed by imposing that behaviour, obtaining
     \be\label{gMode}
     h_k(t)=\frac{e^{-\frac{\lambda \pi}{2}}}{\sqrt{m}} W_{-i\lambda,\frac14}(i\bar t).
     \ee
     These modes define the late-times adiabatic vacuum. As the adiabatic vacuum state constructed for $a(t)=a_0t^p$ with $p>0$, it is a state of infinite adiabatic order at any time. 
     Although the late times adiabatic vacuum is 
      a well justified choice for defining particles at late times, it is obviously not the natural choice as an initial vacuum state. 
      For this reason, one should propose an alternative  construction of the vacuum state at early times.

     \subsection{Early-times conformal symmetry and the vacuum choice}

     An important property of the radiation dominated universe is that the scalar curvature $R$ is identically zero $R=0$. This implies that the scalar field modes propagating in this universe obey the same equation than a conformaly coupled field $\xi=\frac{1}{6}$. One can exploit this property to define a natural vacuum state at early times. \\

    To construct this vacuum, it is useful to expand the solution to Eq. \eqref{rad-h} in terms of a different basis: the $M_{\kappa, \mu}(z)$ Whittaker functions  \cite{mat-functions},
     \be
     \label{solutionM}
      h_{k}=m^{-1/ 2}\Big[B_{k} M_{-i \lambda, \frac{1}{4}}(i \bar t\,)+C_{k} M_{i \lambda, -\frac{1}{4}}(-i \bar t\,)\Big].
     \ee
   
     Again, $B_k$ and $C_k$ are dimensionless constants that have to be settled conveniently. In order to verify the Wronskian condition \eqref{Wronskian}, they have to satisfy the constraint $(B_kC^*_k+C_k B^*_k)=-2$. 
     We now propose a possible way to fix $B_k$ and $C_k$. Firstly, we note that at early times ($t\to 0$), the mass becomes irrelevant in Eq. \eqref{rad-h}, and the mode equation tends to
     \be \label{asymptotic-h}
     \frac{d^{2} h_{k}}{d t^{2}}+\left(\frac{k^{2}}{a_{0}^{2} t}+\frac{3}{16 t^{2}}\right) h_{k}\simeq 0.
     \ee
    This is precisely the equation of a conformal massless scalar field, which admits the exact solutions 
    
    \be h_k\sim \frac{1}{\sqrt{\omega_k}}e^{\mp i\int^t\omega(t')dt'}=t^{1/4}\sqrt{\frac{a_0}{k}}e^{\mp 2i\frac{kt^{1/2}}{a_0}} \ . \ee 
    
    Therefore, in this case we have a natural choice for the initial condition of \eqref{solutionM}. We  select the positive-frequency solution to define the vacuum state. With this early-times behaviour, we are implicitly fixing the coefficients $B_k$ and $C_k$.
    We can expand the solution  \eqref{solutionM} around $\bar t=0=t$ 
     \bea
     h_k\sim m^{-1/2} \Big[C_k(-i\bar t\,\,)^{1/4}+B_k(i \bar t\,\,)^{3/4}+ \cdots\Big] \ ,
     \eea
     while, expanding the positive-frequency conformal solution around $t=0$
     \be
     h_k \sim \sqrt{\frac{a_0}{k}}t^{1/4}-2i\sqrt{\frac{k}{a_0}}t^{3/4}+ \cdots  \ .\label{early-h}
     \ee
     Requiring the matching of both expansions we can fix the vacuum state
     \be \label{early-times-coeffs}
     B_k=-\sqrt{2}(i \lambda)^{1/4},\qquad C_k=\frac{1}{\sqrt{2}}\frac{1}{( -i \lambda)^{1/4}}.
     \ee
     One can check that the Wronskian condition \eqref{Wronskian} is satisfied with these fixed coefficients. We will refer this choice of the vacuum as the {\it early-times non-adiabatic vacuum state}.  This vacuum agrees with the one obtained in Ref. \cite{particle-creation-76} by other means.
     However, the very crucial point is to see whether this vacuum  is consistent with renormalization. In order to answer this question  we investigate the finiteness 
     of the renormalized stress-energy tensor in the following subsection. In Appendix \ref{ap:particle-production} we describe the late-times particle production associated with this vacuum.

\subsection{Renormalizability of the (non-adiabatic) initial vacuum state}

We compute the regularized vacuum expectation values of the two-point function $\langle \phi^2 \rangle $ and the stress-energy tensor $\langle T_{\mu \nu} \rangle $ and study its convergence for the vacuum state \eqref{solutionM}. As we have already explained, an unavoidable requirement that a suitable vacuum state must satisfy is that it has to be UV regular. In the context of cosmological FLRW spacetimes, this condition is ensured by the adiabatic regularity condition, that guarantees the existence of a finite (renormalizable) vacuum expectation value for the two-point function $\langle \phi^2 \rangle$ and the stress-energy tensor $\langle T_{\mu \nu} \rangle $. 
However, we will see that the usual adiabatic regularity condition can be soften in some special cases. \\

In homogeneous and isotropic spacetimes where the adiabatic regularization prescription can be applied, we say that an observable $\langle O\rangle_{\textrm{ren}}\sim \int dk\ k^2 [O(k,t)-O(k,t)_{\textrm{Ad}}]$ 
associated with a particular quantum state is {\it renormalizable} if its momentum integral is finite in the ultraviolet regime.  
 As stressed before, to guarantee the convergence of the 
 renormalized stress-energy tensor $\langle T_{\mu\nu} \rangle_{\textrm{ren}}$, we  usually need to ensure that the (divergent) large frequency behaviour of the modes  coincides with the large frequency behaviour of the adiabatic subtractions up to $4th$ order (in terms of $|h_k|^2$ it means coincidence up to and including $\mathcal{O}\left(k^{-5}\right)$). 
 We note that, for the radiation dominated universe, where $R=0$, there are some simplifications in the adiabatic terms. For the two-point function it results into the following large-momentum expansion ($k\to \infty$)
\bea
 (W_{k}^{-1})^{(0)}&\sim&\frac{a_0\sqrt{t}}{k}-\frac{m^2a_0^3t^{3/2}}{2k^3}+\frac{3m^4a_0^5 t^{5/2}}{8k^5}-\frac{5 m^6 a_0^7 t^{7/2}}{16 k^7}+ \mathcal{O}(k^{-9})\label{WAd0}\\
 (W_{k}^{-1})^{(2)}&\sim& 
 \frac{m^2 a_0^5  t^{1/2}}{16k^5}-\frac{5 m^4a_0^7t^{3/2}}{16 k^7}+\mathcal{O}(k^{-9})\label{WAd2}\\
 (W_{k}^{-1})^{(4)}&\sim&
 \mathcal{O}(k^{-9})\label{WAd4} \  .
 \eea
 
On the other hand, we can perform the large momentum expansion of $|h_k|^2$ for the {\it early-times non-adiabatic vacuum} \eqref{early-times-coeffs}
\bea\label{hkexpansion1}
|h_k|^2 \sim &&\frac{a_0  \sqrt{t}}{k}-\frac{a_0^3 m^2 t^{3/2}}{2 k^3}+\frac{3 a_0^5 m^4 t^{5/2}}{8 k^5}+\frac{a_0^5 m^2 \sqrt{t}}{16 k^5}-\frac{5 a_0^7 m^6 t^{7/2}}{16 k^7}-\frac{5 a_0^7 m^4 t^{3/2}}{16 k^7}\\
&&-\frac{a_0^5 m^2 \sqrt{t} \cos
   \left(\frac{4 k \sqrt{t}}{a_0}\right)}{16 k^5} +\frac{a_0^6
   m^4 t^2 \sin \left(\frac{4 k \sqrt{t}}{a_0}\right)}{24 k^6}+\frac{a^7 m^6 t^{7/2} \cos \left(\frac{4 k \sqrt{t}}{a_0}\right)}{72
   k^7}+\frac{a_0^7 m^4 t^{3/2} \cos \left(\frac{4 k \sqrt{t}}{a_0}\right)}{32 k^7} \nonumber \\
   &&+ \mathcal{O}\left(k^{-9}\right) \  . \nonumber 
\eea
This expression has to be compared with the expansion to large $k$ of the adiabatic terms up to fourth order \eqref{WAd0}, \eqref{WAd2} and \eqref{WAd4}. We can observe in \eqref{hkexpansion1} that the divergent terms for the two-point function (up to $k^{-3}$ in $|h_k|^2$ ) are canceled by the adiabatic subtractions of \eqref{WAd0}. However, a new oscillating term decaying as $k^{-5}$ appear in the expansion for large $k$, and therefore this vacuum does not strictly obey the {\it adiabatic regularity condition}. Nevertheless these oscillatory terms are crucially finite when integrated in \eqref{TPFfinalN}, therefore the two-point function for this vacuum is consistent with renormalization. \\

Although the two-point function is a very direct observable where one can have some insight about the renormalizability of a given vacuum state, the most relevant quantity  is the  stress-energy tensor. For simplicity, and without loss of generality, we fix here the conformal parameter to be $\xi=0$. The formal vacuum expectation values of the energy density $\langle \rho \rangle=\langle T_{00} \rangle$ and the pressure $\langle p \rangle=\frac{1}{3}\sum_{i=1}^3\langle  T_{ii} \rangle$ are given by the following expressions
\bea \label{VEVDensity}
\langle \rho \rangle=&&\frac{1}{4\pi^2 a^3}\int_{0}^{\infty}dk k^2 \frac{1}{2}\Big(|\dot h_k -\frac{3}{2}\frac{\dot a}{a}h_k|^2 + \omega_k^2|h_k|^2 \Big)\equiv \frac{1}{4\pi^2 a^3}\int_{0}^{\infty}dk k^2 \rho(k,t)\, ,\\
\nonumber \\
\langle p \rangle=&&\frac{1}{4\pi^2 a}\int_0^\infty dk k^2\frac{1}{2}\Big(|\dot h_k-\frac{3}{2}\frac{\dot a}{a}h_k|^2 -\frac{1}{3}(\omega_k^2+2m^2)|h_k|^2\Big)\equiv \frac{1}{4\pi^2 a}\int_0^\infty dk k^2 p(k,t)\, , \label{VEVPressure}
\eea
and the trace of the stress-energy tensor is given by $\langle T^{\,\mu}_{\mu} \rangle \equiv\langle T \rangle= \langle \rho \rangle - \frac{3}{a^2}\langle p \rangle$. These quantities are ultraviolet divergent, and have also to be renormalized via the adiabatic method. 
For the stress-energy tensor, the adiabatic subtractions have to be performed up to and including the $4th$ adiabatic order, namely
\bea
\rho(k,t)_{\textrm{ren}}=\rho(k,t)-\rho(k,t)^{(0)}_{\textrm{Ad}}-\rho(k,t)^{(2)}_{\textrm{Ad}}-\rho(k,t)^{(4)}_{\textrm{Ad}}\,,\label{DensityRen}\\
\nonumber\\
p(k,t)_{\textrm{ren}}=p(k,t)-p(k,t)^{(0)}_{\textrm{Ad}}-p(k,t)^{(2)}_{\textrm{Ad}}-p(k,t)^{(4)}_{\textrm{Ad}}\,.\label{PressureRen}
\eea
The explicit expressions of the adiabatic subtractions can be found in Appendix \ref{ap:subtractions}.\\

\begin{figure}[h]
   \includegraphics[width=0.5\textwidth]{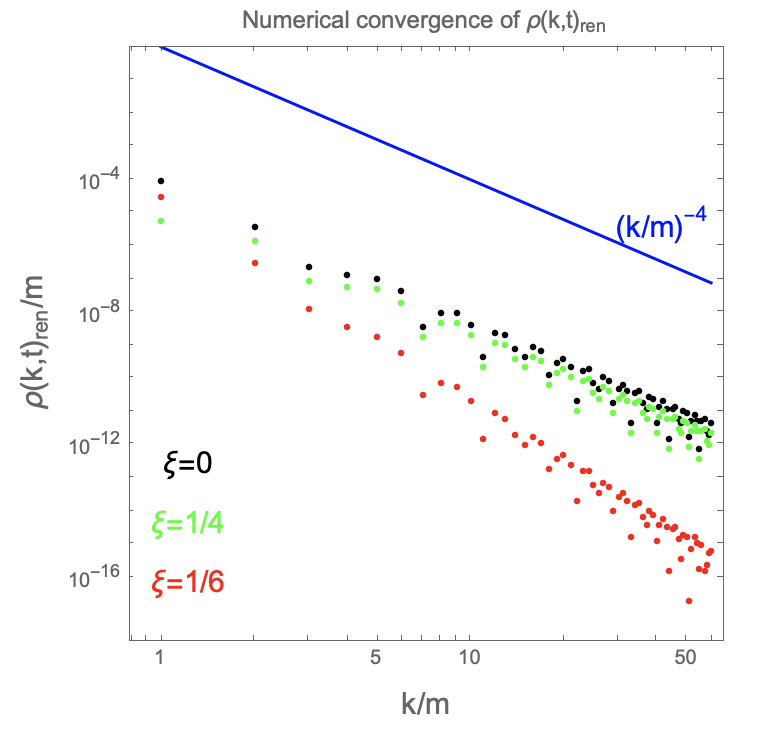}
    \caption{{\small Double logarithmic plot for $\rho(k,t)$ for $a_0/\sqrt{m} =1$, and $mt=1$. Black dots represent the case $\xi = 0$. Green dots represents $\xi = 1/4$. Red dots represent $\xi= 1/6$. The blue line represents $1/k^4$}.}
    \label{T00Convergence}
\end{figure}

We have  checked analytically for $\xi = 0$, that the large momentum expansion of the energy density contains the same divergent terms as the adiabatic expansion. For ${k \to \infty}$ we find
\be \label{rho-largek}
\rho(k,t) \sim \frac{k }{a_0 \sqrt{t}} +\frac{a_0 m^2 \sqrt{t}}{2 k}+\frac{a_0 }{8 k t^{3/2}}-\frac{a^3_0 m^4 t^{3/2}}{8
   k^3}+\frac{a^3_0 m^2}{16 k^3 \sqrt{t}} -\frac{a^4_0 m^{2} \sqrt{m t} \sin \left(\frac{4 k \sqrt{t}}{a_0}\right)}{32 k^4 t^{3/2}}+\mathcal{O}\left( k^{-5} \right)
\ , \ee
while the adiabatic expansion of the energy density for large $k$ reads
\be
\rho(k,t)_{\rm Ad}^{(0)}+\rho(k,t)_{\rm Ad}^{(2)}+\rho(k,t)_{\rm Ad}^{(4)}  \sim \frac{k }{a_0 \sqrt{t}} +\frac{a_0 m^2 \sqrt{t}}{2 k}+\frac{a_0 }{8 k t^{3/2}}-\frac{a^3_0 m^4 t^{3/2}}{8
   k^3}+\frac{a^3_0 m^2}{16 k^3 \sqrt{t}}+\mathcal{O}\left( k^{-5} \right)\, .
\ee
As for the two-point function, we see that we can renormalize the energy density, although finite oscillatory terms remain when performing the subtractions.
We also compute the large momentum expansion for the pressure density $p(k,t)$ 

\bea \label{p-largek}
p(k,t) \sim &&\,\frac{1}{3} a_0 k \sqrt{t} +\frac{a^3_0 }{8 k
   \sqrt{t}} -\frac{a^3_0 m^2 t^{3/2}}{6 k} +\frac{a^5_0 m^4 t^{5/2}}{8 k^3}+\frac{a^5_0 m^2 \sqrt{t}}{48 k^3}\\
   && +\frac{a^5_0 m^2 \sqrt{t}
   \cos \left(\frac{4 k \sqrt{t}}{a_0}\right)}{24 k^3}-\frac{a^6_0 m^4 t^2 \sin \left(\frac{4 k \sqrt{t}}{a_0}\right)}{36 k^4}-\frac{a^6_0 m^2 \sin \left(\frac{4 k
   \sqrt{t}}{a_0}\right)}{32 k^4} + \mathcal{O}\left( k^{-5}\right)\, . \nonumber
\eea
If we compare this expression with the adiabatic expansion of $p(k,t)$ for large $k$,

\be
p(k,t)_{\rm Ad}^{(0)}+p(k,t)_{\rm Ad}^{(2)}+p(k,t)_{\rm Ad}^{(4)}  \sim \frac{1}{3} a_0 k \sqrt{t} +\frac{a^3_0 }{8 k
   \sqrt{t}} -\frac{a^3_0 m^2 t^{3/2}}{6 k} +\frac{a^5_0 m^4 t^{5/2}}{8 k^3}+\frac{a^5_0 m^2 \sqrt{t}}{48 k^3}  + \mathcal{O}\left( k^{-5}\right) \ ,
\ee
we see that the divergences of $p(k,t)$ are canceled by the adiabatic terms. It is important to point out that oscillatory terms like $\cos{\left(c k \right)}/ k^3$ are UV regular.\footnote{ This is somewhat similar to the behaviour of a continuum integral of oscillating terms in the Riemann-Lebesgue lemma of Fourier analysis. This behavior is crucial in perturbative quantum field theory \cite{Tong-lectures}. } We can conclude that the stress-energy tensor for this vacuum choice is {\it renormalizable } although it does not strictly satisfy the adiabatic regularity condition. Therefore, the early-times vacuum is a legitimate vacuum choice with initial conditions 
near the $t=0$  singularity. Finally, one can check 
that the stress-energy tensor is sensitive to the coupling $\xi$ of the scalar field with the curvature $R$. Figure \ref{T00Convergence} shows that the energy density is renormalizable for different values of $\xi$. Note that for the conformal case $\xi=\frac{1}{6}$, the momentum integral of the energy density converges more rapidly.

\section{Conclusions and final comments}
\label{sec:conclusions}

We have discussed the problem of selecting a preferred vacuum state in FLRW spacetimes for scalar fields. It is well known that, except for very specific cases, such as Minkowski or De Sitter space-times, there is not a preferred choice. However, there are two important conditions that any suitable vacuum has to meet: it has to respect the symmetries of the space-time, and it has to be UV regular. These two conditions reduce the space of possible solutions, but there is still an inherent ambiguity that one has to manage. In adiabatic regions of the (FLRW) spacetime, (e.g., $t\to 0$ for $a(t)=t^p$ with $p>1$ or $t\to \infty$ for $a(t)=t^p$ with $p<1$) one can naturally single out a preferred solution 
 which is, by construction, UV regular. \\

However, when fixing the vacuum in a non-adiabatic region of the spacetime one has to find another criterion to fix it  in a consistent way. In this work we have focused in a  radiation-dominated universe. For a scalar field  the modes become conformal when $t\to 0$ [see Eq. \eqref{asymptotic-h}], and therefore a natural solution can be favoured  [see Eq. \eqref{early-times-coeffs}]. It clearly differs from the vacuum constructed in adiabatic regions.  
By this reason it is unclear that the selected vacuum  is consistent with renormalization. We have analyzed with detail this issue. 
We have found that, although the proposed vacuum does not strictly satisfy the usual  adiabatic regularity condition, it turns out to be UV regular in a broader sense [see Eqs. \eqref{hkexpansion1},  \eqref{rho-largek} and \eqref{p-largek}]. \\

Finally, we would like to note that the proposed non-adiabatic vacuum for the radiation-dominated universe  is also CPT invariant in the analytic continuation of $a(\tau) \propto \tau$ to negative values of the conformal time $\tau$. It is not difficult to check that the modes (\ref{solutionM}, \ref{early-times-coeffs}), when properly re-expresed in terms of the parabolic cylinder functions, obey the CPT-invariant condition $h_k(-\tau) = h_k^*(\tau)$. This could be relevant for the CPT-symmetric universe proposed in \cite{turok18, turok21}. We further analyze this issue in \cite{Beltran-CPT}.

\section*{Acknowledgements}
  
 S.P. thanks E. Winstanley and the Gravitation and Cosmology Research Group (CRAG) for their hospitality during her visit to the University of Sheffield, where part of this work was carried out. S.N. also expresses his gratitude to I. Agulló and the LSU Department of Physics and Astronomy for their warm reception during his visit to Louisiana State University, where part of this work was carried out. 
 This work is supported by the Spanish Grant 
  PID2020-116567GB-C21 funded by MCIN/AEI/10.13039/501100011033, and the project PROMETEO/2020/079 (Generalitat Valenciana). 
P. B. is supported by the Ministerio de Ciencia, Innovaci\'on y Universidades, Ph.D. fellowship, Grant No. FPU17/03712. S. N. is supported by the Universidad de Valencia, within the Atracci\'o de Talent Ph.D fellowship No. UV-INV- 506 PREDOC19F1-1005367. S. P. is supported by the Ministerio de Ciencia, Innovaci\'on y Universidades, Ph.D. fellowship, Grant No. FPU16/05287.

\appendix
\section{Particle production for the early-times non-adiabatic vacuum}
\label{ap:particle-production}

     For the early times conformal vacuum of the radiation-dominated universe \eqref{solutionM}, it is possible to study the particle production at late times. To this end, we find it useful to re-write the mode function $h_k(t)$ in terms of the $W_{\kappa,\mu}(z)$ basis, that is 
     \bea
   h_k(t)&=&\frac{1}{m^{1/2}}\Big[B_k M_{-i \lambda, \frac{1}{4}}(i \bar t\,)+C_k M_{i \lambda, -\frac{1}{4}}(-i \bar t\,)\Big] \label{hkprev}\\
   &=& \frac{e^{-\frac{\lambda \pi}{2}}}{m^{1/2}}\Big[\alpha_k W_{-i \lambda, \frac{1}{4}}(i \bar t\,)+\beta_k W_{i \lambda, \frac{1}{4}}(-i \bar t\,)\Big]\, ,
     \eea
     with $B_k$ and $C_k$ given in Eq. \eqref{early-times-coeffs} and where we have extracted the factor $e^{-\frac{\lambda \pi}{2}}$ for future convenience. The new coefficients $\alpha_k$ and $\beta_k$ satisfy the condition $|\alpha_k|^2-|\beta_k|^2=1$. These two different basis are related by \cite{mat-functions}
   {\small \bea \label{M14-W}
    M_{-i \lambda,\frac{1}{4}}(i\bar t\,)&=&e^{-\pi \lambda}\Bigg[\frac{i^{3/2}}{\Gamma(\frac{3}{4}-i \lambda)}W_{-i\lambda,\frac{1}{4}}(i\bar t\,) + \frac{1}{\Gamma(\frac{3}{4}+i \lambda)}W_{i\lambda,\frac{1}{4}}(-i\bar t\,)\Bigg]\frac{\sqrt{\pi}}{2} \,,\\
   M_{i \lambda,-\frac{1}{4}}(-i\bar t\,)&=&e^{- \pi\lambda}\Bigg[ \frac{1}{\Gamma(\frac{1}{4}-i \lambda)}W_{-i\lambda,\frac{1}{4}}(i\bar t\,)+\frac{i^{1/2}}{\Gamma(\frac{1}{4}+i \lambda)}W_{i\lambda,\frac{1}{4}}(-i\bar t\,) \Bigg]\sqrt{\pi}\, .
    \label{M14-Wb}\eea}
    Implementing \eqref{M14-W} and \eqref{M14-Wb} into \eqref{hkprev} we immediately find $\alpha_k$ and $\beta_k$. From this result, the analysis of the particle production is straightforward. As we have stressed before, the radiation dominated universe becomes more and more adiabatic as $t\to \infty$. It means that, at late times, the solution will be a linear combination of positive and negative frequency solutions. Indeed, using the asymptotic behaviour ($t\to \infty$) of the Whittaker $W$ functions \cite{mat-functions} we find
    \bea
    h_k(t)\sim \frac{1}{\sqrt{m}}(\alpha_k e^{-\frac{i \bar t\,}{2}}(\bar t\,)^{-i \lambda} + \beta_k e^{\frac{i \bar t}{2}}(\bar t\,)^{i \lambda}).
    \eea
    As usual \cite{parker-toms}, a non-vanishing $\beta_k$ coefficient accounts for particle production. In particular, it can be shown that the  average density number of created particles $n_k$, with momentum $k$, is given by the coefficient $|\beta_k|^2$, that is 
    \be \label{particle-num}
    n_k=|\beta_k|^2=e^{-\pi \lambda}\frac{ \pi }{2}\Bigg(\frac{\lambda^{-1/2}}{|\Gamma(\frac{1}{4}+i\lambda)|^2}+\frac{\lambda^{1/2}}{|\Gamma(\frac{3}{4}+i\lambda)|^2}\Bigg)-\frac{1}{2}.
    \ee
    It is interesting to analyze the large $k$ behaviour of the particle number density. The large $k$ asymptotic expansion ($k\to\infty$) of Eq. \eqref{particle-num} gives
    \be
    |\beta_k|^2 \sim\frac{1}{16384\lambda^4} + O(\lambda^{-6})
    \ee
   in agreement with \cite{particle-creation-76}.

\section{Adiabatic subtractions for the stress-energy tensor with $\xi=0$} \label{ap:subtractions}

 With the adiabatic expansion of the mode function (\ref{adiabatic-expansionN}), we can compute the adiabatic expansions of our main observables, and subtract them to their formal (unrenormalized) values 
up to a given order to obtain a finite and meaningful result. For the quantities derived from the stress-energy tensor we have to make subtractions up to and including the 4th adiabatic order. The results used in the main text are 

\bea
\rho(k,t)_{\textrm{Ad}}^{(0-4)}&=&\omega _k+\frac{m^4 \dot{a}^2}{8 a^2 \omega _k^5}+\frac{m^2 \dot{a}^2}{2 a^2 \omega
   _k^3}+\frac{\dot{a}^2}{2 a^2 \omega _k}-\frac{105 m^8 \dot{a}^4}{128 a^4 \omega
   _k^{11}}-\frac{21 m^6 \dot{a}^4}{16 a^4 \omega _k^9}+\frac{33 m^4 \dot{a}^4}{16 a^4
   \omega _k^7}+\frac{3 m^2 \dot{a}^4}{4 a^4 \omega _k^5}+\frac{3 \dot{a}^4}{8 a^4
   \omega _k^3} \\
   &+&\frac{7 m^6 \dot{a}^2
   \ddot{a}}{16 a^3 \omega _k^9}+\frac{15 m^4 \dot{a}^2 \ddot{a}}{16 a^3 \omega
   _k^7}-\frac{\dot{a}^2 \ddot{a}}{4 a^3 \omega _k^3}+\frac{m^4 \ddot{a}^2}{32 a^2 \omega _k^7}+\frac{m^2 \ddot{a}^2}{8 a^2 \omega
   _k^5}+\frac{\ddot{a}^2}{8 a^2 \omega _k^3} -\frac{m^4 \dot{a} a^{(3)}}{16 a^2 \omega _k^7}-\frac{m^2 \dot{a}
   a^{(3)}}{4 a^2 \omega _k^5}-\frac{\dot{a} a^{(3)}}{4 a^2 \omega _k^3}\, , \nonumber 
\eea

\bea
p(k,t)_{\textrm{Ad}}^{(0-4)}&=&-\frac{m^2}{3 \omega _k}+\frac{\omega _k}{3}+\frac{5 m^6 \dot{a}^2}{24 a^2 \omega
   _k^7}+\frac{3 m^4 \dot{a}^2}{8 a^2 \omega _k^5}+\frac{\dot{a}^2}{6 a^2 \omega
   _k}-\frac{385 m^{10} \dot{a}^4}{128 a^4 \omega _k^{13}}-\frac{259 m^8 \dot{a}^4}{128
   a^4 \omega _k^{11}}+\frac{7 m^6 \dot{a}^4}{a^4 \omega _k^9}
   -\frac{13 m^4 \dot{a}^4}{16
   a^4 \omega _k^7}\\
   &+&\frac{m^2 \dot{a}^4}{8 a^4 \omega _k^5}
   +\frac{\dot{a}^4}{8 a^4 \omega
   _k^3}-\frac{m^4
   \ddot{a}}{12 a \omega _k^5} 
   -\frac{m^2 \ddot{a}}{3 a \omega _k^3}-\frac{\ddot{a}}{3 a
   \omega _k}+\frac{77 m^8 \dot{a}^2 \ddot{a}}{32 a^3 \omega _k^{11}}+\frac{49 m^6 \dot{a}^2
   \ddot{a}}{16 a^3 \omega _k^9}-\frac{4 m^4 \dot{a}^2 \ddot{a}}{a^3 \omega _k^7}
   -\frac{5
   m^2 \dot{a}^2 \ddot{a}}{4 a^3 \omega _k^5}\nonumber \\
   &-&\frac{\dot{a}^2 \ddot{a}}{2 a^3 \omega
   _k^3}
   -\frac{7 m^6 \ddot{a}^2}{32 a^2 \omega _k^9}-\frac{15 m^4 \ddot{a}^2}{32 a^2
   \omega _k^7}+\frac{\ddot{a}^2}{8 a^2 \omega _k^3} 
   -\frac{7
   m^6 \dot{a} a^{(3)}}{24 a^2 \omega _k^9}-\frac{5 m^4 \dot{a} a^{(3)}}{8 a^2
   \omega _k^7}+\frac{\dot{a} a^{(3)}}{6 a^2 \omega _k^3}\nonumber \\
   &+&\frac{m^4 a^{(4)}}{48 a
   \omega _k^7}+\frac{m^2 a^{(4)}}{12 a \omega _k^5}+\frac{a^{(4)}}{12 a \omega
   _k^3}\, .\nonumber 
\eea

\end{document}